\RequirePackage{fix-cm}
\documentclass[smallextended]{svjour3}       
\smartqed  
\usepackage{graphicx}

\usepackage{amsmath}
\usepackage[authoryear]{natbib}
\usepackage[ot2enc]{inputenc}
\usepackage[russian,english]{babel}
\usepackage{longtable}
\usepackage{epstopdf}

 \journalname{}
\begin{document}

\title{A data-based classification of Slavic languages: Indices of qualitative variation applied to grapheme frequencies\thanks{M. Ko\v s\v cov\'a and J. Ma\v cutek were supported by VEGA grant 2/0047/15.}
}

\titlerunning{A data-based classification of Slavic languages}        

\author{Michaela Ko\v s\v cov\'a  \and  J\'an Ma\v cutek \and Emmerich Kelih
}

\authorrunning{Michaela Ko\v s\v cov\'a et al.} 

\institute{M. Ko\v s\v cov\'a \at
              Department of Applied Mathematics and Statistics, Comenius University, Mlynsk\'a dolina, SK-84248 Bratislava, Slovakia\\
           \and
           J. Ma\v cutek \at
              Department of Applied Mathematics and Statistics, Comenius University, Mlynsk\'a dolina, SK-84248 Bratislava, Slovakia\\
              Tel.: +421-2-60295717\\
              Fax: +421-2-65412305\\
              \email{jmacutek@yahoo.com}
           \and
			E. Kelih \at
			Department of Slavonic Studies, University of Vienna, Spitalgasse 2, Hof 3, AT-1090 Wien, Austria
}

\date{Received: date / Accepted: date}

\maketitle

\begin{abstract}
The Ord's graph is a simple graphical method for displaying frequency distributions of data or theoretical distributions in the two-dimensional plane. Its coordinates are proportions of the first three moments, either empirical or theoretical ones. A modification of the Ord's graph based on proportions of indices of qualitative variation is presented. Such a modification makes the graph applicable also to data of categorical character. In addition, the indices are normalized with values between 0 and 1, which enables comparing data files divided into different numbers of categories. Both the original and the new graph are used to display grapheme frequencies in eleven Slavic languages. As the original Ord's graph requires an assignment of numbers to the categories, graphemes were ordered decreasingly according to their frequencies. Data were taken from parallel corpora, i.e., we work with grapheme frequencies from a Russian novel and its translations to ten other Slavic languages. Then, cluster analysis is applied to the graph coordinates. While the original graph yields results which are not linguistically interpretable, the modification reveals meaningful relations among the languages.
\end{abstract}

\section{Introduction and motivation}
\label{sec:intro}

\begin{figure}

\centering

\setlength{\unitlength}{1cm}

\begin{picture}(10,11.5)
\put(0.5,0){\vector(0,1){11.5}}
\put(0,5.5){\vector(1,0){6.5}}
\put(0.5,0.5){\line(1,2){5.7}}
\put(0.5,10.5){\line(1,0){5}}
\put(5.5,5.5){\line(0,1){5}}
\put(6.5,5.2){\scriptsize{$I$}}
\put(0.3,11.5){\scriptsize{$S$}}
\put(0.3,5.2){\scriptsize0}
\put(0.3,10.4){\scriptsize1}
\put(0.2,0.5){\scriptsize{-1}}
\put(5.4,5.2){\scriptsize1}
\put(0.5,10.5){\circle*{0.1}} 
\put(0.6,10.6){\scriptsize{$A$}}
\put(5.5,10.5){\circle*{0.1}} 
\put(5.6,10.3){\scriptsize{$P$}}
\put(0.5,0.5){\circle*{0.1}} 
\put(0.6,0.3){\scriptsize{$G$}}
\put(3,5.5){\circle*{0.1}} 
\put(3,5.2){\scriptsize{$J$}}
\put(6,11.2){\scriptsize{$B$}}
\put(6.2,11.5){\vector(1,2){0.18}}
\put(2.2,3.5){\scriptsize{Binomial (line segment GP)}}
\put(1.4,8.5){\scriptsize{Hypergeometric}}
\put(1.0,8.2){\scriptsize{(area of triangle AGP)}}
\put(1.7,11.3){\scriptsize{Beta-Pascal (above AP,}}
\put(1.9,11){\scriptsize{left of half-line PB)}}
\put(6,7){\scriptsize{Beta-binomial (half-plane below GP)}}
\put(6.5,11){\scriptsize{Negative binomial (half-line PB)}}
\put(5,2){\scriptsize{point J = Symmetric binomial}}
\put(5,1.7){\scriptsize{point P = Poisson}}

\end{picture}

\caption{Graphical representation of discrete distributions from \cite{Ord:1967b}.\label{fig:obr_ord}}
\end{figure}
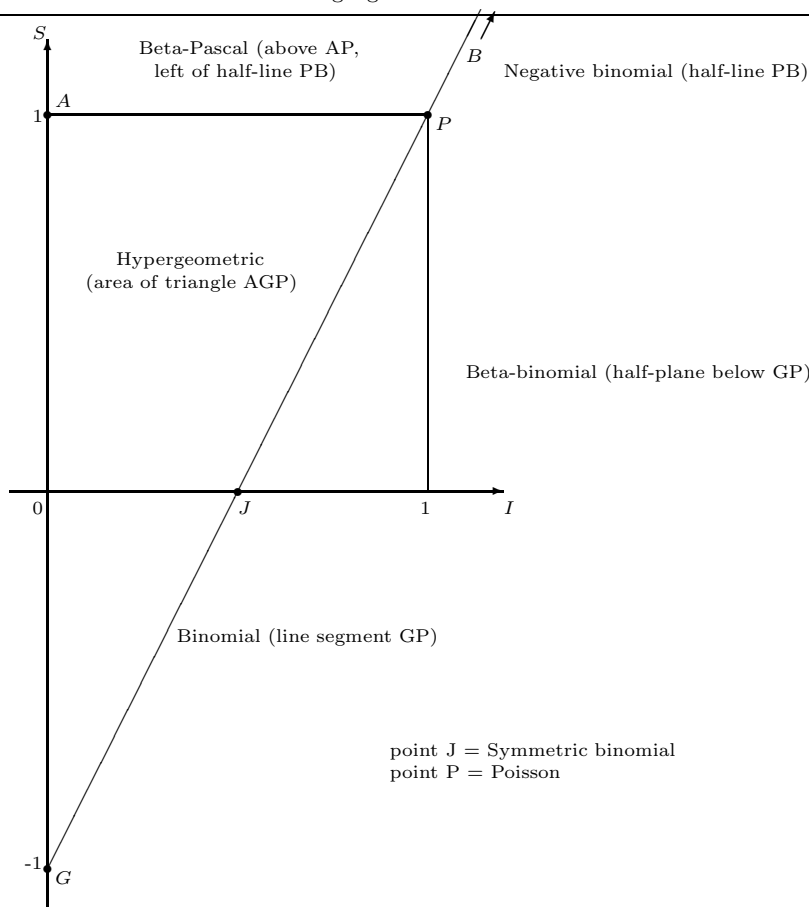

\cite{Ord:1967b} suggested a simple graphical representation of discrete probability distributions \footnote{In order to avoid confusion, we remind that the same author also developed another graphical method for discrete distributions, which was published in the same year, see \citealp{Ord:1967a}, and also \citealp{Friendly:2000}} in the two-dimensional plane -- however, his idea can directly be applied  also to continuous distributions. The coordinates of a distribution in the graph are given as proportions of their first three moments, namely, the mean $\mu$, the variance $\mu_2$ and the third central moment (i.e., the skewness) $\mu_3$. In general, all distribution can be depicted for which the first three moments exist and the first two of them are non-zero. Keeping the notation from \cite{Ord:1967b}, the $x$- and $y$-coordinates will be denoted by $I$ and $S$, respectively, with $I=\mu_2 / \mu $ and $S=\mu_3 / \mu_2$. If all possible parameter values of a particular distribution are considered, one obtains an area (or a curve, a line, a point) characteristic for the distribution (we note that areas belonging to different distributions can overlap). Some of them can be seen in Figure~\ref{fig:obr_ord}, which is taken from \cite{Ord:1967b}.

If theoretical moments are replaced with empirical ones, the Ord's graph can also be used to display data. It can serve as a preliminary, intuitive decision criterion whether the data can be modelled by a particular distribution. If the point representing the data lies within the area of the distribution, or not too far away from it, a (relatively) good fit between the data and the model can be expected. The graph provides also, among others, a possibility of data classification or clustering -- points representing related data are supposed to be close to each other.

The graph was applied to visualize data in quite distinct branches of research, like, e.g., biology \citep*{Schneider_Duffy:1985}, transport networks modelling \citep*{Taylor:1976, Beguin_Thomas:1997}, linguistics \citep*{Stadlober_Dzuzelic:2005, Grzybek_Rusko:2009}, and musicology \citep{Martinakova_et_al:2009}.

However, the Ord's graph is not applicable to data of categorical character (for an overview of graphical methods suitable for such data see \citealp{Blasius_Greenacre:1998}, and \citealp{Friendly:2000}). Especially in the case of nominal data, i.e., if there is no natural ordering of categories (see, e.g., \citealp[p.~3]{Agresti:2013}), using the graph would require an assignment of integers to the categories. Such an assignment can only be arbitrary, and the arbitrariness leads almost necessarily to ambiguities.

We will apply both the original Ord's graph and its new modification (see Section~\ref{sec:indices}) to grapheme frequencies in Slavic languages (see Section~\ref{sec:data} for data description). Grapheme orderings, as they are established in alphabets (or other writing systems) specific for particular languages, are results of traditions and/or conventions which are not linguistically substantiated in the vast majority of languages. Slavic languages are not exceptional in this respect. Moreover, two further facts mar any attempt to achieve a grapheme ordering common to all Slavic languages. They not only have different grapheme inventories, but languages from this family also use writing systems based on two different scripts, namely, Latin and Cyrillic. These two scripts (and their modifications) follow different traditions of grapheme orderings, e.g., the grapheme {\it z} appears towards the end of Slavic adaptations of the Latin alphabet (\citealp{Comrie:1996b}), but its Cyrillic counterpart {\it\cyr z} is positioned around the eighth place (out of roughly 30, depending on the language, see Section~\ref{sec:data}) in alphabets based on the Cyrillic script (\citealp{Comrie:1996a}). 

One of reasonable possibilities left is to work with ranked frequencies, where the most frequent grapheme is given the rank 1, the second most frequent the rank 2, etc. The problem of ambiguities mentioned above is thus solved. This direction of research enjoyed an increased popularity in recent years. There are several studies available, mainly for Slavic languages (see \citealp{Grzybek_et_al:2009}, and references therein), but also for German (\citealp{Grzybek:2007}), Irish and Manx (\citealp{Wilson:2013}), and some languages from West Africa (\citealp{Rovenchak_Vydrin:2010}). The negative hypergeometric distribution (see, e.g., \citealp{Wimmer_Altmann:1999}, pp. 465--468) is tentatively considered a general mathematical model. However, its parameters, and hence also its moments, seem to depend on the inventory size (i.e., on the number of graphemes used in particular languages, IS henceforward; the determination of the grapheme inventory size is a complex linguistic issue, some details specific for Slavic languages can be found in \citealp{Kelih:2013}). The dependence within the Slavic language family was demonstrated by \cite{Grzybek_Kelih:2005} and \cite{Grzybek_et_al:2005}. Consequently, also the Ord's graph, which exploits the moments, will reflect not only a measure of relatedness among Slavic languages, but it will also be influenced by their inventory sizes. We will show in Section~\ref{sec:data} that the graph constructed from grapheme rank-frequency distributions does not lead to linguistically explainable results. 

Therefore, in Section~\ref{sec:indices} we suggest a modification of the Ord's graph, in which moments are replaced with so-called indices of qualitative variation (see \citealp{Wilcox:1973}). The new graph reveals a meaningful classification of Slavic languages.

\section{Data description}\label{sec:data}

The grapheme frequencies which will be analyzed were obtained from the Russian social realist novel {\it Kak zakaljalas' stal'} ({\it How the Steel Was Tempered}) and its translations to ten other Slavic languages. The book was written by Nikolai Ostrovsky in 1930s. It enjoyed the status of recommended reading; therefore it was translated to the languages spoken in the countries from the socialist bloc within a relatively short time period. The linguistic corpus consisting of the Russian (RUS henceforward, $\textrm {IS}=33$) original and its translations into Belorusian, Bulgarian (BUL, $\textrm {IS}=30$), Croatian (CRO, $\textrm {IS}=30$), Czech (CZE, $\textrm {IS}=42$), Macedonian (MAC, $\textrm {IS}=31$), Polish (POL, $\textrm {IS}=32$), Serbian (SRB, $\textrm {IS}=30$), Slovene (SLO, $\textrm {IS}=25$), Slovak (SVK, $\textrm {IS}=43$), Ukrainian (UKR, $\textrm {IS}=34$), and Upper Sorbian (UPS, $\textrm {IS}=37$) was described by \cite{Kelih:2009b}.

\begin{figure}
\begin{center}
\includegraphics[scale=0.4]{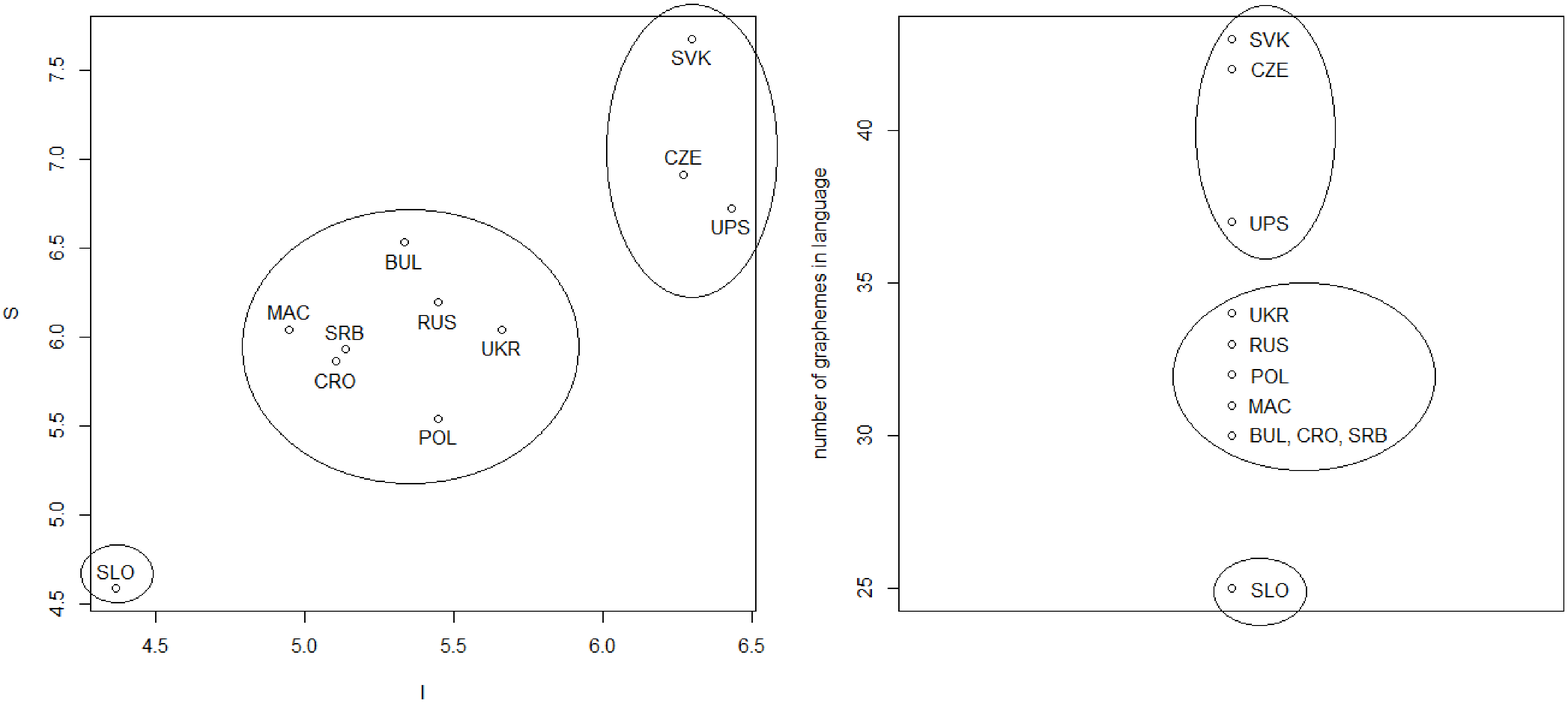}
\end{center}
\caption{Original Ord's graph applied to grapheme frequencies (left) and inventory sizes in Slavic languages (right), with cluster analysis applied to graph coordinates.} \label{fig:original_ord_and_number_graphemes}
\end{figure}

Belorusian was omitted from our considerations, as its orthography differs substantially from other Slavic languages. Belorusian has an explicit, phonetically determined orthographic system, i.e., letters are used for coding phones and not phonemes (and partly morphophonemes) as, e.g., in case of Russian and Ukrainian. This different coding approach has, among others, the effect of an extreme overexploitation of particular graphemes (for details see \citealp{Kelih:2009a}). Rank-frequency distributions of graphemes from eleven Slavic languages can be found in Table~\ref{table:grapheme_frequencies} (the languages are ordered decreasingly according to their grapheme inventory sizes); they are displayed on the Ord's graph in Figure~\ref{fig:original_ord_and_number_graphemes} left. \footnote{Two from among currently spoken standard Slavic languages were not included: Belorusian, as was explained, was omitted because of its peculiar orthography; and Lower Sorbian, because no suitable texts (i.e., long enough and comparable with analogous texts in other Slavic languages) could be found (the language has about 7000 speakers only). We do not intend to discuss here the status of one language/different languages/dialects of, e.g., Ukrainian/Rusyn, Bosnian/Croatian/Montenegrin/Serbian, Polish/Cassubian, etc.}

\begin{longtable}{cccccccccccc}
\caption{Grapheme rank-frequency distributions in Slavic languages.}
\endfirsthead
\caption[]{continued}
\endhead
\label{table:grapheme_frequencies}																			
	&	SVK	&	CZE	&	UPS	&	UKR	&	RUS	&	POL	&	MAC	&	BUL	&	CRO	&	SRB	&	SLO	\\	
\hline
1	&	26490	&	20618	&	29440	&	25494	&	28305	&	26718	&	40232	&	36841	&	32444	&	32507	&	30849	\\	
2	&	23869	&	20371	&	27097	&	22419	&	23509	&	25264	&	30122	&	24724	&	25820	&	25823	&	29708	\\	
3	&	20564	&	19595	&	24691	&	17958	&	21205	&	22229	&	28420	&	23098	&	24952	&	24709	&	26129	\\	
4	&	15166	&	15223	&	17213	&	16868	&	17140	&	20509	&	20985	&	21644	&	24320	&	23473	&	25886	\\	
5	&	13204	&	14183	&	16201	&	15985	&	16143	&	18622	&	20793	&	19535	&	13457	&	13332	&	17175	\\	
6	&	12842	&	12586	&	14719	&	14123	&	14868	&	14275	&	17111	&	17133	&	13215	&	13168	&	15921	\\	
7	&	12233	&	12174	&	13527	&	12146	&	13980	&	13344	&	13634	&	13867	&	12958	&	12888	&	15045	\\	
8	&	12137	&	11365	&	12224	&	11835	&	13265	&	12876	&	13152	&	13394	&	12759	&	12728	&	14144	\\	
9	&	11959	&	11312	&	11500	&	11566	&	13103	&	12627	&	11613	&	12224	&	11581	&	11453	&	14139	\\	
10	&	11548	&	10193	&	10995	&	10521	&	12693	&	12120	&	10640	&	11329	&	10237	&	9949	&	12402	\\	
11	&	10010	&	9639	&	10640	&	10339	&	10004	&	11170	&	10591	&	9197	&	9958	&	9929	&	11569	\\	
12	&	8981	&	9147	&	10113	&	9926	&	8396	&	10120	&	7815	&	8542	&	9885	&	9661	&	11412	\\	
13	&	8569	&	8477	&	9647	&	9811	&	8147	&	9637	&	7753	&	7950	&	9741	&	9163	&	10029	\\	
14	&	8293	&	8320	&	8425	&	8871	&	7834	&	9499	&	7123	&	7339	&	9139	&	8296	&	9167	\\	
15	&	7389	&	8252	&	7725	&	8327	&	7733	&	8933	&	6327	&	6197	&	8384	&	7958	&	8753	\\	
16	&	6729	&	6301	&	7697	&	7542	&	5479	&	8623	&	6127	&	5633	&	7779	&	7794	&	6441	\\	
17	&	6051	&	5552	&	7238	&	6693	&	5191	&	8510	&	5440	&	5309	&	5047	&	5015	&	5515	\\	
18	&	5496	&	5338	&	7182	&	5640	&	5045	&	6564	&	5219	&	4770	&	4688	&	4732	&	5336	\\	
19	&	4282	&	5229	&	5625	&	4759	&	5026	&	5964	&	5191	&	4554	&	3808	&	3889	&	4755	\\	
20	&	4270	&	5219	&	5540	&	4618	&	4957	&	5354	&	4360	&	4344	&	3768	&	3797	&	4429	\\	
21	&	4267	&	4719	&	5341	&	4215	&	4498	&	4613	&	3203	&	4035	&	3075	&	3004	&	3054	\\	
22	&	3697	&	4207	&	5201	&	3977	&	3679	&	4387	&	2015	&	3220	&	2258	&	2239	&	2923	\\	
23	&	3352	&	4103	&	4135	&	3952	&	3288	&	4361	&	1798	&	2681	&	2225	&	2194	&	1967	\\	
24	&	2772	&	3290	&	4024	&	3038	&	2859	&	3714	&	1540	&	2197	&	1810	&	1832	&	1893	\\	
25	&	2498	&	3169	&	3579	&	2963	&	2667	&	3199	&	803	&	1956	&	1769	&	1703	&	230	\\	
26	&	2424	&	2932	&	3412	&	2486	&	2506	&	2548	&	563	&	1936	&	1709	&	1592	&		\\	
27	&	2358	&	2650	&	2888	&	2101	&	1556	&	2052	&	365	&	1464	&	1665	&	1512	&		\\	
28	&	1867	&	2583	&	2867	&	1937	&	1098	&	1851	&	303	&	362	&	637	&	649	&		\\	
29	&	1722	&	2460	&	2813	&	1430	&	971	&	1220	&	171	&	336	&	241	&	278	&		\\	
30	&	1456	&	2098	&	2668	&	1340	&	539	&	416	&	66	&	320	&	55	&	77	&		\\	
31	&	1276	&	2032	&	2241	&	878	&	312	&	406	&	35	&		&		&		&		\\	
32	&	642	&	892	&	607	&	282	&	59	&	254	&		&		&		&		&		\\	
33	&	601	&	541	&	505	&	242	&	0	&		&		&		&		&		&		\\	
34	&	581	&	253	&	276	&	1	&		&		&		&		&		&		&		\\	
35	&	366	&	213	&	0	&		&		&		&		&		&		&		&		\\	
36	&	320	&	188	&	0	&		&		&		&		&		&		&		&		\\	
37	&	186	&	182	&	0	&		&		&		&		&		&		&		&		\\	
38	&	100	&	169	&		&		&		&		&		&		&		&		&		\\	
39	&	94	&	86	&		&		&		&		&		&		&		&		&		\\	
40	&	30	&	12	&		&		&		&		&		&		&		&		&		\\	
41	&	10	&	7	&		&		&		&		&		&		&		&		&		\\	
42	&	6	&	0	&		&		&		&		&		&		&		&		&		\\	
43	&	0	&		&		&		&		&		&		&		&		&		&		\\	
\hline
\end{longtable}

Since the beginning of modern Slavic linguistics and typology in the mid-19th century, the classification of Slavic languages has been discussed many times. By now, a simple typology based on the geographical location of the Slavic standard languages is more or less accepted; it divides the languages into three groups: East Slavic (Belorusian, Russian, Ukrainian), West Slavic (Czech, Polish, Slovak, Upper and Lower Sorbian), and South Slavic (Bulgarian,  Croatian, Macedonian, Serbian, Slovene). 

Cluster analysis was applied to the $I$- and $S$-coordinates from the Ord's graph, with three clusters required (indicated by ellipses in Figure~\ref{fig:original_ord_and_number_graphemes} left). Clustering was performed in statistical software R. Two methods were used, namely, k-means and k-medoids. In Figure~\ref{fig:original_ord_and_number_graphemes} left, they yield the same clusters regardless of the choice of the algorithm for the k-means method (Hartigan-Wong, Lloyd, MacQueen) and of the metric for the k-medoids method (Euclidean, Manhattan). Figure~\ref{fig:original_ord_and_number_graphemes} right presents clusters resulting of the k-means method; the k-medoids method gives clusters almost identical to the ones from Figure~\ref{fig:modified_ord} (the only difference is that UPS migrates into the cluster containing CZE and SVK). For a (relatively) short overview of the cluster analysis see, e.g., \cite{Izenman:2008}. 

The results obtained are not linguistically meaningful (e.g., East Slavic languages form one group with most of South Slavic ones; on the other hand, Slovene is a single outlier, which is not explainable, since the historical development of its writing system is parallel with the other Slavic languages, etc.). The only clue hinting towards a linguistic explanation is the grapheme inventory size of the languages analyzed, as the clusters coincide with the ones based on the sizes of grapheme inventories (Figure~\ref{fig:original_ord_and_number_graphemes} right). Grapheme inventories, however, reflect history, traditions, conventions, etc. of a language (see also Section~\ref{sec:intro}) more than linguistic laws and relations among languages; furthermore, they are extremely conservative and almost resistant to changes (which, if occur, follow more often than not sudden historical/political changes, and not slow, continuous changes of languages).

Given that moments of the grapheme rank-frequency distributions depend, at least for Slavic languages, on the inventory sizes (\citealp{Grzybek_Kelih:2005}; \citealp{Grzybek_et_al:2005}), the coincidence of clusters in Figure~\ref{fig:original_ord_and_number_graphemes} is not surprising.

\section{Modified Ord's graph}\label{sec:indices}

Consider $N$ data items divided into $K$ categories and denote $f_i$ the frequency of the $i$-th category. \cite{Wilcox:1973} discussed in his paper several measures of variation applicable (also) to nominal data, among them the variance analogue
\begin{equation}\label{eq:va}
VA = 1-\frac{\sum_{i=1}^{K}\left(f_i - \frac{N}{K}\right)^2}{\frac{N^2 \left(K-1\right)}{K}},
\end{equation}
the standard deviation analogue
\begin{equation}\label{eq:sda}
SDA = 1-\sqrt{\frac{\sum_{i=1}^{K}\left(f_i - \frac{N}{K}\right)^2}{\frac{N^2 \left(K-1\right)}{K}}},
\end{equation}
and the relativized entropy
\begin{equation}\label{eq:re}
RE = \frac{-\sum_{i=1}^{K}p_i \log p_i}{log K},
\end{equation}
where $log$ denotes the natural logarithm.

In \cite{Wilcox:1973}, these measures are called indices of qualitative variation. They have at least two properties which distinguish them from the usual measures of variation (like the variance, the standard deviation, etc.). First, they are invariant with respect to the ordering of categories, i.e., they depend solely on frequencies. Second, all of them are normalized, with possible values from the interval $[0,1]$ (for all of them, value $0$ is attained if all objects are in one category and other categories are empty; value 1 corresponds to the uniform distribution, with all categories having the same frequencies). Thus, if one considers grapheme frequencies in Slavic languages, indices of qualitative variation can be a response to ambiguities related to the two traditions of grapheme orderings. They also eliminate influences of different inventory sizes.

Given these advantages, we applied the indices (\ref{eq:va})-(\ref{eq:re}) to modify the Ord's graph. The modified coordinates are defined as 
\begin{equation}\label{eq:modified_I}
I_m = SDA / VA
\end{equation}
 and 
\begin{equation}\label{eq:modified_S}
S_m = RE / SDA. 
\end{equation}

It is easy to see that $I_m$ could be simplified to the form 
\begin{equation*}
I_m = 1+\sqrt{\frac{\sum_{i=1}^{K}\left(f_i - \frac{N}{K}\right)^2}{\frac{N^2 \left(K-1\right)}{K}}}.
\end{equation*}
However, out of two reasons we prefer here to keep the form (\ref{eq:modified_I}); first, to highlight an analogy with the original Ord's graph (other measures of qualitative variation can be more useful for analyses of other types of data, see Section~\ref{sec:conclusion}). Second, specifically for linguistic data, the form (\ref{eq:modified_I}) can be more simple to interpret. Its denominator is, in fact, the normalized repeat rate 
\begin{equation*}
RR_{norm} = \frac{K}{K-1}\left(1-\frac{\sum_{i=1}^{K}f_i^2}{N^2}\right),
\end{equation*}
see \cite{Gibbs_Poston:1975}, which is one of the standard characteristics in linguistics.

\begin{figure}
\begin{center}
\includegraphics[scale=0.48]{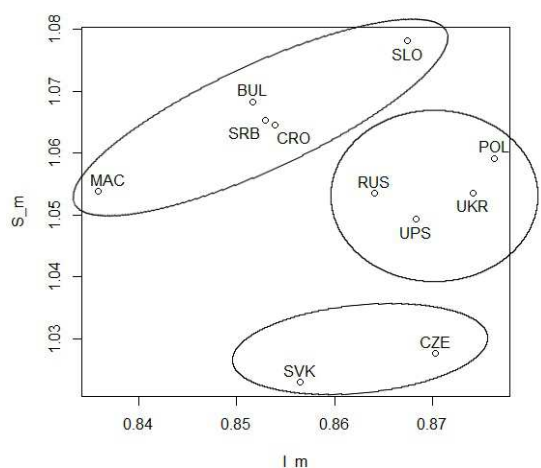}
\end{center}
\caption{Modified Ord's graph applied to grapheme frequencies, with cluster analysis applied to graph coordinates.}\label{fig:modified_ord}
\end{figure}

In Figure~\ref{fig:modified_ord} the new graph can be seen, applied, again, to grapheme frequencies in Slavic languages (we emphasize that the order of graph\-emes within a language is irrelevant in this case). Clusters created from its coordinates $I_m$ and $S_m$ (ellipses in Figure~\ref{fig:modified_ord}) present a pattern quite different from Figure~\ref{fig:original_ord_and_number_graphemes}. The proposed classification reveals interesting findings on the typology of Slavic languages (the resulting clusters are the same, again, regardless of the method, algorithm or metric used, see Section~\ref{sec:data}).

First of all, there is a group of South Slavic languages, which perfectly fits with their geographical location. Bulgarian, Croatian, Macedonian, Serbian, and Slovene form one homogenous group. The orthographic systems of these languages are well organized with respect to the economy of coding of some specific prosodic features (like the pitch accent in Croatian, Serbian, and Slovene) and to the marking of palatalized consonants in Bulgarian (marked with a specific vocalic grapheme). Macedonian is one of the youngest standard languages (codified in 1945) and its orthography is largely based on the same principles as Serbian (one letter for one sound).

The second group can be called the basic West Slavic languages, it includes Czech and Slovak. The two languages are typologically quite similar in general, including their orthographic and phonemic systems, and thus their location in one group is justified. 

In Figure~\ref{fig:modified_ord}, Russian, Ukrainian, Polish and Upper Sorbian form one group. If one compares it with the traditional geographical classification, this North Slavic group seems to be a mixture of East Slavic (Russian and Ukrai\-nian) and West Slavic languages (Polish and Upper Sorbian). However, if orthographic and phonemic criteria are taken into account, these languages share some common features, namely, they are characterized by a systematic correlation of the consonantal system palatalization (i.e., consonants tend to have both ``hard'' and ``soft'' versions). Indeed, these characteristics play a very important role in Russian and Ukrainian, whereas a regression of palatalization was reported for Polish and especially for Upper Sorbian.

The groups resulting from the cluster analysis of the modified Ord's graph coordinates differ slightly from the traditional, area-based typology of Slavic languages, but they suggest another, linguistically justifiable classification. It corresponds to the approach of \cite{Kolomiec:1986}, where a group of North Slavic languages (Russian, Ukrainian, Polish) is mentioned; they are characterized by a high number of consonants in their inventories, whereas South Slavic languages mainly enlarged their vowel inventory (for a detailed discussion of vocalic and consonantal Slavic languages see \citealp{Sawicka:1991}).

\section{Conclusion}\label{sec:conclusion}

Our modification of the Ord's graph brings linguistically motivated and interpretable results. Cluster analysis applied to the coordinates of the new graph reveals groups of languages which share some common features, as far as orthography and phonology is concerned. Thus, the application of the modified Ord's graph to grapheme frequencies can be seen as a contribution towards the typology of Slavic languages. If compared with their traditional, purely geographical classification, the new approach has the advantage of being based on empirically observed data.

Admittedly, the definition of the modified graph coordinates (\ref{eq:modified_I}) and (\ref{eq:modified_S}) used in this paper -- i.e., the choice of indices (\ref{eq:va})-(\ref{eq:re}) -- is heuristic only. Apart from the fact that they yield linguistically relevant results in this case, there is no other reason why they should be preferred. It can be expected that other indices of qualitative variation (see, e.g., \citealp{Wilcox:1973}; \citealp{Gibbs_Poston:1975}) can be more reasonable for categorical data arising from other branches of science. 

Regardless of the choice of the indices, the method is computationally very simple; results it yields are also easy to understand, as they are displayed in the two-dimensional plane. In addition, it represents categorical data by two real-valued coordinates, enabling thus applications of statistical classification or clustering methods.


\bibliographystyle{spbasic}       
\bibliography{qualitative_variation_04_14}   

%
%

\end{document}